# Unsupervised learning of part similarity for goal-guided accelerated experiment design in metal additive manufacturing

Rui Liu[1,‡,*], Sen Liu[2,‡,*], and Xiaoli Zhang[3,4]

[1] Cognitive Robotics and AI Lab, College of Aeronautics and Engineering, Kent State University, Kent, OH 44240, USA

[2] Department of Mechanical Engineering, University of Louisiana at Lafayette, Lafayette, LA 70503, USA

[3] Department of Mechanical Engineering, Colorado School of Mines, Golden, CO 80401, USA

[4] Alliance for the Development of Additive Processing Technologies, Colorado School of Mines, Golden, CO 80401, USA

‡ Equally contributed.

* Correspondence author; E-mails: sen.liu@louisiana.edu; rliu11@kent.edu.

**Abstract:** Metal additive manufacturing (AM) is gaining broad interest and increasing use in the industrial and academic fields. However, the quantification and commercialization of standard parts usually require extensive experiments and expensive post-characterization, which impedes the rapid development and adaptation of metal AM technologies. In this work, a similarity-based acceleration (S-acceleration) method for design of experiments is developed using K-means clustering and Gaussian Mixture Model (GMM) unsupervised learning algorithms to reduce the time and costs associated with unveiling process-property (porosity defects) relationships during manufacturing. With S-acceleration, part semantic features from machine-setting parameters and physics-effects informed characteristics are explored for measuring mutual part similarities. A user-defined simplification rate of experiments is proposed to purposely remove redundant parts before conducting experiments printing without sacrificing information gain as original full factorial experiment design. This S-acceleration design of experiments is demonstrated on a Concept Laser M2 Dual-laser Cusing machine for the experimental plan of modeling relationships between process parameters and part porosity defects. The printed part has 2 mm diameter × 4 mm tall pin geometry considering variations in build location and orientation, laser settings and powder feedstock are held constant. In total, 242 parts are measured to create a "ground truth" data set of porosity levels by using X-ray tomography microscopy. The S-acceleration method is assessed for performance considering 40%, 50%, and 60% of user-defined experiment simplification rates. The repeated experiments are removed without ignoring the minority experiments outlier, assuring a similar process-property relation in the original experiment plan. The experiment number is significantly reduced based on part similarity with minimal compromise of model accuracy and obtained knowledge.

# 1. Introduction

Metals additive manufacturing (AM) technologies are gaining interest and adoption in the manufacturing industry for building customized products such as jet engines and body shells, automotive products, and medical devices [1-3]. Laser powder bed fusion (LPBF) is a widely used metal AM technology that produces parts with a complex thermal history that originates from directional solidification followed by multiple heating and cooling cycles as the laser passes over subsequent layers. Because of the short laser-material interaction times and highly localized heat input, the thermal gradients and rapid cooling rate lead to a build-up residual stress, cracks, non-equilibrium microstructures, geometric distortion, mechanical properties, and defects such as lack-of-fusion and key-hole porosity [4-8]. The current open research remains including the efficient experiment design before printing, physics-informed model insight for defects prediction, process parameter optimization, and in-process quality control [5,9-11,13-14,53-57].

Traditionally, the quantification and certification of AM part property are performed after materials deposition. As a result, most of the product development follows trial and error manufacturing and testing for comprehensive quality assurance, which causes intensive labor and expensive resources. Metals AM experiments and post-processing characterization are expensive due to factors such as the high cost of making quality metallic feedstocks, especially alloy powders, the cost of acquiring and operating AM machines, and property characterizations such as X-ray computed microtomography for assessing porosity microstructure and mechanical properties [14-18,50]. For example, in this study, to elucidate the relationships of machine-setting parameters (*e.g.*, laser power, laser speed, layer thickness, part location and orientation on the plate, *etc.*) and printed part porosity defects, there are 605 parts printed on each build plate for different machine-setting parameter variations. About 3 hours are needed to characterize the part porosity defect with X-ray tomography microscope. The entire cost for data collection exceeds $120,000 and 75 days in an academic program; double this cost in the industry. Even with the state-of-the-art characterization technologies, the costs and time are intractable to qualify the basic process-property (PP) relationships. Therefore, it is critical to develop a strategy to simplify the experiment plan in the experiment design stage before actually doing experiments; that is, to reduce the number of parts needed for PP relationships establishment, at the same time without sacrificing the information gain as traditional full factorial experiments design.

The advancement of the data-driven machine learning (ML) method has been shown to accelerate the adaptation of new processes and property design by modeling and predicting for part internal defects, surface quality, geometry distortion, microstructures, and mechanical property [19-23,52-57]. For example, A process map was constructed for powder bed fusion AM using a support vector machine to classify built parts with good or bad quality [22]. A model was developed to construct maps of molten pool depth or porosity level versus process parameters for parts fabricated by LPBF AM [24,25]. Aboutaleb et al. [26] have accelerated process optimization for laser-based additive manufacturing by leveraging similar prior studies. They leveraged three papers with 22 experiments datasets as initial experiments of design. It then predicts the property at new design points by maximizing the response function (that is, model exploitation). Finally, the model is updated by adding a unique design point. Through five times iteration of experiments, it sequentially identifies a high-density part > 99% in the real-world case study. Liu et al. [5,12] demonstrated a metal AM learning and process optimization framework to accelerate the adoption of new printing technologies by reusing the previously acquired knowledge of manufacturing conditions, fabricated part properties from multiple printers. The framework is verified through three "industry-use" inspired scenarios for laser powder bed fusion of Ti-6Al-4V. Khanzadeh et al. [51] predict locations of porosity almost 96 % of the time during directed energy deposition processes by utilizing an appropriate Self-Organizing Maps (SOMs) model and capturing the melt pool signals. Liu et al. [5,12], perform feature extraction of online high-dimensional senor data of additive manufacturing. The online process monitoring methodology based on supervised classification and regression models identifies actual process quality status.

Despite these data-driven ML approaches and experimental studies, the concept of experiment simplification methodology to reduce the number of experiments needed to deduce process-property relationships knowledge of metal AM is rarely seen in the literature. Researchers traditionally used Design of experiments (DoE) methodology such as Taguchi orthogonal array to design the process parameters combinations [27-31]. For example, Griffiths et al. [29] utilized a full factorial Taguchi L16 orthogonal array to define four process parameters combinations, and 320 experimental trials in total being carried out to correlate the output responses such as energy consumption and production time. Braconnier et al. [30] defined the DoE parameter space (*e.g.*, extrusion temperature, layer thickness, printed bed temperature, print speed, and printer type) to study the influences of process parameters on print strength across three commercial printers. Manjunath et al. [31] also studied the effect of print parameters such as laser travel speed, accelerating voltage, and beam current on deposited layer width was analyzed using the Taguchi method and ANOVA feature analysis. Although the Taguchi and full factorial DoE methods can significantly reduce the experimental trials and economic cost, these methods have been criticized in the literature for some disadvantages in accounting for interactions between process parameters and the mechanistic insight gain is limited [32-34]. These methods rely purely on pre-defined feature levels and orthogonal array, ignoring the intrinsic physics characteristics during manufacturing. Insightful descriptions of the experiment plan are lacking for quantifying the similarities of process parameters combination. Moreover, with the increasing number of processing parameter variations (*e.g.* over tens of dimensional spaces), the traditional full factorial experiments design is challenging to adapt without a guideline to define the threshold or similarity level to quantify the produced part similarities. A more advanced experiment design methodology for quantification of the similarity and clustering underlying physics mechanism of each process combination in the original experiments candidates is becoming essential to reduce redundant parts printing and conduct a most informative and economic experiment plan.

In this work, we adopt a feature similarity-based acceleration (S-acceleration) method to realize the goal of reducing the number of experiments needed for establishing process-property (porosity defects) relationships in LPBF AM. Specifically, to measure mutual part similarity, the S-acceleration method uses part semantic features related to machine-setting parameters, part positions on the plate and physics-effects characteristics during the laser-materials interaction. In this manner, the framework is possible to apply and generalize to different machines, materials and processing conditions where the simplification is independent of specific machine settings and parameters. The similar parts in the experiments design space are grouped together with semantic features, and the repetitive parts in each group of experiments plan are intentionally removed before real-world practice printing to reduce the number of parts printing and subsequent characterization process. This method is inspired by the "Facebook friend recommendation" system for social networks [35] in which two persons are highly likely to be friends if they share intrinsic characteristics such as education and interests, and outer characteristics such as social circles and vice versa. Our contributions in this paper are two-fold:

(1) A novel similarity measurement algorithm was proposed to measure printed parts' mutual repetitiveness aiming at a simplified experiment plan by printing lesser samples needed for learning knowledge of process-property relationships. Semantic aspects/features, such as physical effects, machine setting parameters, and porosity defects, are investigated as intrinsic and extrinsic features to describe the printed part. The part similarities are measured by spatial distance between the parts in the designed semantic feature space. The similarities and differences among parts provide theoretical foundations to assess part similarity in property characteristics such as mean and maximum pore size.

(2) A novel goal-guided experiment simplification method based on simplification rate was developed to reduce the number of parts printed to draw the equivalent process-property relationships as traditional full factorial experiments design. Specifically, the goal defines simplification intensity and focuses on maintaining data probability distribution to achieve research conclusions similar to those of the plan of the original experiment. During S-acceleration DoE simplification, similar parts are clustered together; highly repeated parts are removed from the clusters with different proportions and similarity level strategies.

## 2. Methods

### 2.1. Goal definition in AM modeling

The machine-setting parameters such as laser power and laser speed involve a full of process parameters variations, fluctuation, and respective interactions. Under a specific manufacturing condition, the parts present a spatial pattern distribution with different intrinsic characteristics and performance. It is possible to cluster parts according to their similarities, which are represented as spatial characteristic pattern distributions. Given that both part characteristics and performance are influenced by the same manufacturing conditions, it is reasonable to assume that these parts share similar output properties. Therefore, highly similar parts in a cluster will be considered as highly likely repeated parts. In the original experiments design for different process parameter combinations, by removing these repeated parts, the number of experiment trials will be reduced to simplify the original experiment plan.

To reduce the number of experiments while minimally compromising statistical certainty in the desired determination of PP relationships, two questions should be answered: (1) Which data metric is best to measure part similarity considering the desired knowledge of PP relationships? (2) What is an acceptable trade-off threshold between the number of experiments to perform and the certainty of the PP relationships?

To answer the above questions, the goal of simplification should be identified. It is generally defined as the simplification rate of original experiment candidates to derive the required process-property relationship of the manufacturing process. For example, in this work, the goal is the experiments simplification rate of 0.4, 0.5, 0.6 as user-defined to identify the probability of the maximum pore size in a part given the process parameters such as the position and orientation of the part within the build volume, fixing feedstock and processing parameters. Then, given the goal, metrics for potential semantic part aspects are selected to measure experiment similarities within a planned design of experiments. In the example followed in this work, these metrics obviously include the PP measures of interest: expected distribution of maximum pore sizes, part orientations, and part locations in the build volume, but these metrics are augmented with other physically measurable quantities such as the number of build layers in a part and the analytically approximated laser energy density on each build layer. Next, unsupervised learning algorithms cluster parts based on similarities in these metrics; in this work, K-means clustering is used. Parts within each cluster are ranked from high to low in mutual similarity with respect to the other parts within the cluster. Finally, according to a user-defined percentage, the most redundant experiments are removed from the experiment plan. A schematic of this experiment simplification workflow is shown in Figure 1.

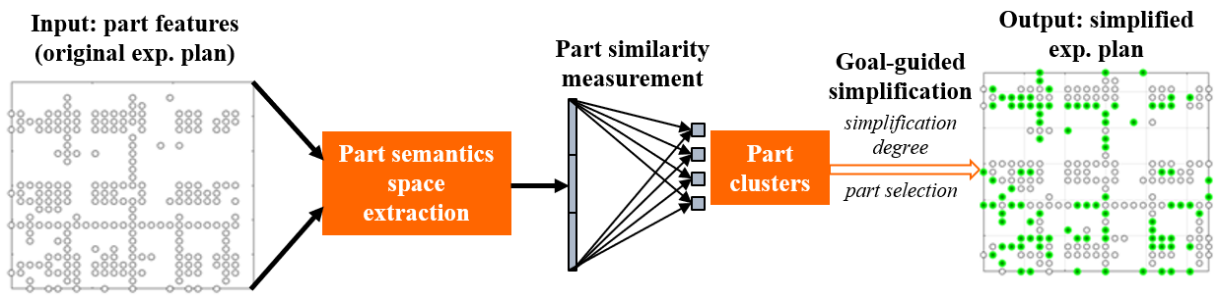

**Figure 1.** A schematic workflow for experiment simplification based on the S-acceleration method. The part characteristics/features are extracted with machine setting parameters and various physical effects. By analyzing part similarity, parts are grouped into different part clusters using an unsupervised learning method. Parts in each cluster are selected with goal-guided simplification degrees for a simplified experiment plan.

*2.2. Semantics-based part similarity measurement*

According to experiment goal G, potential goal-related semantic aspects/features are selected as $s_1$, $s_2$, ... to construct a semantic space. For each aspect, several semantic features are selected to describe the part's semantic characteristics (refer to how an entity exists in space such as shape, size, and material). For example, when the semantic aspect "machine setting" is selected to measure part similarity, the features, such as laser power, part column/row coordinate, part polar angle, and part azimuth angles, are selected to describe a part's machine setting conditions. The selection of goal-related semantic features is based on previous research conclusions or engineering experience.

Given that all the parts are produced under the same process parameters such as laser power and laser speed, the microstructure and mechanical properties also change gradually among parts according to each part's location and angle on the build plate, as shown in build plate in Figure 1. Therefore, besides intrinsic different machine-setting process parameters, neighboring parts' environment reflects part characteristics to some degree. When two parts are similar, both the intrinsic and their neighboring characteristics are similar, and vice versa. It is therefore necessary to measure part mutual similarity by measuring the characteristics of the part themselves and with their neighbor relationships. To formalize our similarity metrics, (1) the similarity aspects of intrinsic part characteristics are measured, and (2) the mutual similarities with part neighbors are calculated by the coordinates of parts in the semantic space. A Euclidean distance between two parts is measured to indicate two parts' mutual similarities. Let $a$ and $b$ denote two random parts, whose similarities are waiting to be measured. Overall, part mutual aspect similarity feature score $s_{aspect}(a,b)$ of part $a$ and $b$ is calculated by part intrinsic similarity feature and neighbor similarity feature, which is expressed as:

$$s_{aspect}(a,b) = u \cdot s_{intrinsic}(a,b) + v \cdot s_{neighbor}(a,b) \quad (1)$$

Where $u$ and $v$ are the contribution weights of similarity factors, calculated according to manufacturing experience. In this paper, $u$ and $v$ were defined as 0.5 with equal importance.

The intrinsic similarity feature is calculated with Eq. (2). Part similarity measurements consider common and non-common features in each semantic aspect of a part. Let $G_a$, $G_b$ denote the feature sets for parts $a$ and $b$, respectively. $G_a \cap G_b$ denotes the common features shared by part $a$ and $b$; $G_a \cup G_b$ denotes the feature integration of part $a$ and $b$; $\Psi(\cdot)$ denotes a weighted linear combination method for similarity calculation [36]. Similarity based on parts' intrinsic characteristics is encouraged by common features $\Psi(G_a \cap G_b)$ and discouraged by non-common features $(G_a \cup G_b - G_a \cap G_b)$. Neighbor similarity $s_{neighbor}$ is calculated using Eq. (3). $i$, $j$ are the indexes of surrounding neighbors of part $a$ and $b$, respectively; $G_i$, $G_j$ denote the feature sets of these neighbor parts. In Eq. (2), the intrinsic similarity is calculated by common features between parts $a$ and $b$, which is calculated by the ratio of overlapped features to the overall integrated features; in equation 3, neighbor similarity is decided by the most similar neighbor pair between the $a$ and $b$.

$$s_{intrinsic}(a,b) = \frac{\Psi(G_a \cap G_b)}{\Psi(G_a \cup G_b)} \quad (2)$$

$$s_{neighbor}(a,b) = \max_{i \in R^a, j \in R^b} \frac{\Psi(G_i \cap G_j)}{\Psi(G_i \cup G_j)} \quad (3)$$

The coordinate of each part in the semantic space is calculated with the similarity scores of the part relative to all other parts. Feature importance is defined by parameter $t$, and is decided by the manufacturing requirements. In this study, we considered all the features with equal importance and set each by an average value of $1/N$. N is the sample size in the building plate, that is the dimension of semantic vector of the part. The $s_i^a$ ($i = 1, ... N$) is the part $a$'s mutual similarity feature with respect to other sample index in $i$. The coordinates $L_a$ and $L_b$ for two parts $a$ and $b$ are generated as Eqs. (4-5).

$$L_a = (t_1 \cdot s_1^a, t_2 \cdot s_2^a, t_3 \cdot s_3^a, \ldots t_N \cdot s_N^a) \quad (4)$$

$$L_b = (t_1 \cdot s_1^b, t_2 \cdot s_2^b, t_3 \cdot s_3^b, \ldots t_N \cdot s_N^b) \quad (5)$$

A Euclidean distance measurement method is formalized as for similarity measurement $E(a,b)$, as shown in Eq. (6). The parts with a shorter distance are relatively more similar.

$$E(a,b) = \|L_a - L_b\| = \sqrt{\sum_{i=1}^{N} t_i^2 (s_i^a - s_i^b)^2}, \quad \sum_{i=1}^{N} t_i = 1 \quad (6)$$

## 2.3. Goal-guided experiment simplification

Based on part mutual similarity, a goal-guided experiment simplification is conducted by grouping parts together, ranking all the parts inside the group by part repeating likelihood, and removing highly repeated parts according to the requirements. To group the relatively similar parts together, unsupervised learning algorithms are adopted to create part clusters based on the part similarity scores [37]. To rank part similarity inside a group, mutual similarities of parts inside a cluster are analyzed [38]. The parts in relatively higher similarity with other parts inside a group indicate highly likely repeated parts. To remove highly repeated parts, the goal-required proportion of the parts is selected from high-repeating likelihood to low-repeating likelihood in the similarity ranking. The goal-required proportion is decided by an actual goal set in an experiment. In the experiment simplification process, "goal-guided" is described as follows: First, identify the research purpose for an experiment design. Through simplification, the research goal should be achieved with similar acceptable performance. Second, select semantic features to measure part similarity according to the research purpose. Different research purposes require different factor considerations in part similarity measurement. For example, the part porosity analysis relies more on machine settings and physical effects during laser-materials interaction, while the part yield strength property analysis relies more on part microstructures [6,39,40,41,42,43]. It is necessary to select different semantic aspects of the part to perform part similarity measurement. Third, according to the simplification degree and research purpose, the number of repeated parts in the original full factorial design space would be removed from each cluster.

2.3.1 Similarity-based unsupervised part clustering

To perform the part similarity-based clustering within semantic feature space, the typical machine learning algorithms k-means clustering [44] and Gaussian Mixture Model clustering [45] are used to cluster the data.

To categorize all the parts into K different groups, the unsupervised learning k-means algorithm automatically cluster parts into different groups according to the mutual part distances in the semantic space. Input is $S \in R^N$, where $N$ denotes the total sample number. Each sample is an $N$ dimension semantic vector, and each dimension is a mutual similarity value with another part. The desired category number is $K$ with part clusters. $\mu_k$ is the mean of points in part clusters $k$. By minimizing the within-cluster sum of squares, the optimal cluster objective is to learn,

$$\arg\min_{k,x} \sum_{j=1}^{K} \sum_{x \in s_i} \|\mathbf{s}-\boldsymbol{\mu}_k\|^2 \qquad (7)$$

Gaussian mixture model (GMM) is a parametric probability density function, representing data by parametric Gaussian density distributions. Part similarity could be described by probability concentrations, and these parts could be clustered with the same level of similarity. GMM is represented as:

$$Q\left(\frac{\mathbf{S}}{\Omega}\right) = \sum_{k=1}^{K} \sigma_k q_k\left(\frac{\mathbf{S}}{\varphi_k}\right) \qquad (8)$$

Where parameter $\Omega=(\sigma_1,\ldots,\sigma_k,\varphi_1,\ldots,\varphi_k)$ with a constraint $\sum_{k=1}^{K}\sigma_k=1$. $\sigma_k$ is the coefficient describing the possibility of data $x$ in cluster $k$. $\varphi_k$ is a distribution descriptor for the Gaussian density. $q_k$ denotes Gaussian density distribution in the cluster $k$. $\mathbf{S}=(s_1,s_2,\ldots,s_N)$ is the similarity vector for each part. To decide the optimal clustering plan, GMM will find the optimal $\Omega$ to maximize the probability distribution $Q\left(\frac{\mathbf{S}}{\Omega}\right)$. Maximum likelihood (ML) is used to solve optimal $\Omega$, as shown by

$$G(\Omega)=\log Q\left(\frac{\mathbf{S}}{\Omega}\right) = \sum_{n=1}^{N} \log(\sum_{k=1}^{K} \sigma_k q_k \frac{s_n}{\Omega}) \qquad (9)$$

2.3.2 Part repeating likelihood ranking

After clustering, part mutual similarity inside each cluster is visualized in heatmaps, as shown in Figure 2(Left). Each row shows the mutual similarity of a part with the others. A part's mean similarity is calculated by averaging all the mutual similarities in a row. Based on the mean similarity, the part repeating likelihoods are ranked, shown in Figure 2(Right). Parts ranked with higher places will be considered as being more similar to others and will be removed from the cluster, and parts ranked in relatively low places will be considered as being less repeated and will be retained.

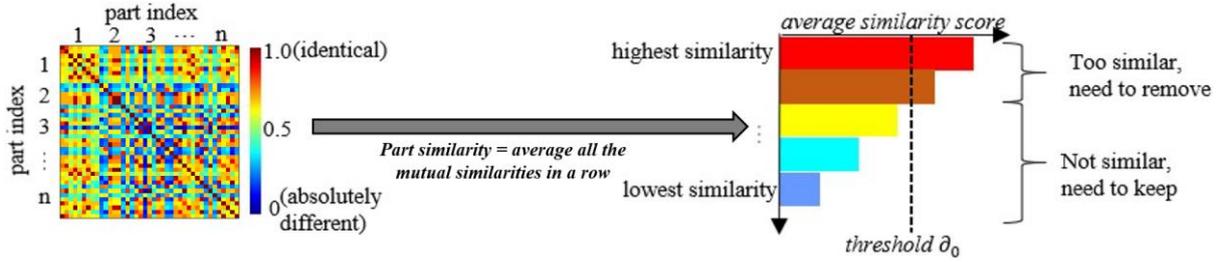

**Figure 2.** Part repeating likelihood analysis based on mutual part similarity. **Left:** a heatmap for visualizing part mutual similarity inside a cluster; **Right:** ranking of part repeating likelihood, which denotes part representative in a cluster. When the part repeating likelihood is high, it will be removed and vice versa.

2.3.3 Part-removing strategy selection for experiment simplification

In this step, the highly repeated parts will be removed to simplify the real-world experiments, and the less repeated parts will be kept to maintain the original part information given the research goal of process-porosity defects relation. A part-removing strategy $\pi$ is designed with the consideration of both the original experiment goal $g$ and the simplification intensity $l$. With the guidance of the part-removing strategy $\pi$, the clusters and their representative parts are selected. A cost function $J_S$ is defined in Eq. (10) to assess the satisfactory degree of the potential experiment simplification strategy. $\pi_{kd}$ denotes the satisfactory score, which is assessed by applying a simplification strategy $\pi$ on part $d$ in cluster $k$, with the consideration of the satisfactory degree of reaching experiment goal $g$ and implementing simplification degree $s$. $\widehat{\pi_{kd}}$ is the average score of $\pi_{kd}$. The experiment strategy $\pi$ with the lowest cost function $J$ will be selected as the final recommended experiment acceleration plan.

$$J_l = \frac{1}{KD}\sum_{k=1}^{K}\sum_{d=1}^{N}(\pi_{kd}(l,g) - \widehat{\pi_{kd}}(l,g))^2 \qquad (10)$$

To validate the effectiveness of the S-acceleration method in experiment simplification, the process – porosity defects relations of LPBF AM research scenario was selected to perform experiment simplification. The simplification plan was determined based on the part characteristics analysis before conducting experiments. For validating the performance of the S-acceleration method on experiment simplification, the original experiment plan was also conducted after experiment simplification. By performing the absolute relative error and standard deviation (std) of simplified degree $s$ relative with the target degree $g$ for maintaining the pore size distribution of the original plan versus the simplified plan, the effectiveness of S-acceleration was validated. The pseudocode structure of S-acceleration for design of experiments workflow is shown in Table 1.

**Table 1.** Pseudocode structure of S-acceleration algorithms.

| **Pseudocode structure of S-acceleration for design of experiments** | |
|---|---|
| 1 | **Input:** experiment simplification goal for simplification rate; semantic aspect parameters such as machine settings, laser status and part geometry and pose. **Output:** the simplified experiments design. |
| 2 | define weights for intrinsic and neighborhood similarities $u, v$; |
| 3 | randomly select two parts a, b from part sample set |
| 4 | initialize $S_{aspect} \leftarrow us_{intrinsic} + vs_{neighbor}$, $u = 0.5$ and $v = 0.5$ with equal importance |
| 5 | calculate each part's sematic features in the coordinate space; $L_a \leftarrow (t_1 \cdot s_1^a, t_2 \cdot s_2^a, t_3 \cdot s_3^a, \ldots t_N \cdot s_N^a)$ $L_b \leftarrow (t_1 \cdot s_1^b, t_2 \cdot s_2^b, t_3 \cdot s_3^b, \ldots t_N \cdot s_N^b)$, $N$ is the dimension of sematic vector of the part, feature weight $t_i = 1/N$ for equal importance. |
| 6 | update $L_a$, $L_b$ for all of the parts in the print plate |
| 7 | calculate similarity clusters of all of the parts in the original experiments design with k-means or GMM. |
| 8 | part repeating likelihood ranking. |
| 9 | part-removing strategy selection for experiment simplification; $J_l \leftarrow \frac{1}{KD}\sum_{k=1}^{K}\sum_{d=1}^{N}(\pi_{kd}(l,g) - \widehat{\pi_{kd}}(l,g))^2$ |
| 10 | evaluation with maximum pore size distribution in histogram plot, absolute error ← \|simplified mean–target simplified degree\| and STD with different simplification methods. |

## 3. Experiments

### 3.1. Machine settings

The metal selective-laser-melting samples were produced on a Concept Laser M2 Dual-Laser Cusing machine. This system has a $27.7 \times 27.7 \times 30$ cm$^3$ build envelope. Inconel 718 powder is spread from the supply chamber over the exposure surface area using the rubber wiper blade. SLM uses an inert gas environment to mitigate oxide and inclusion formation. An active flow of Argon is switched on when the Oxygen level is greater than 0.6%. A 200 W fiber laser is installed 482.60 mm above the laser powder bed surface. The laser beam diameter in the normal direction of laser source is 50 µm. Laser parameters are selected as a linear track pattern at 160 W laser power and 800 mm/s laser speed. Layer thickness is set at 50 µm. The dual laser cusing machine uses two lasers to divide the powder bed exposure surface in half and assign half of the plate area to each laser. The build volume is maintained in an Argon atmosphere with the temperature in the range of 21-27 ºC and relative humidity in the range of 6–18%. To simulate the porosity distribution in different locations of an AM-produced part, the part sample adopted in this experiment is designed to be a small cylinder with a diameter of 2 mm and a height of 4 mm. The machine setting is shown in Figure 3. A plate is visualized as that in Figure 3(a), in which two dots denote the lasers' vertical projections. The build plate dimensions were 247×247 mm. Nine unique orientations of samples are built on the plate. Every row of parts on the plate had a unique tilt angle relative to build plate, as shown in Figure 3(b). A total of 605 samples were built on the plate. Given the time and cost of part inspection, we selected 242 parts without any heat treatments and post machining for porosity analysis in this study. To conduct layer-wise part porosity analysis, a part is simplified as 40 horizontal layers, evenly distributed along the vertical direction with respect to a part's original setting pose, as shown in Figure 3(c).

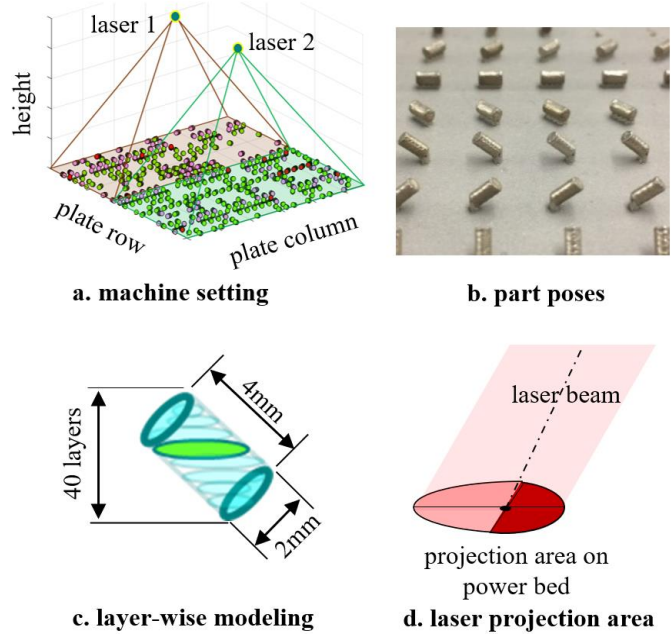

**Figure 3.** **(a)** Two lasers were involved in the manufacturing process. Each laser worked on one-half of the plate. **(b)** Each part on the plate has a unique location and part pose. **(c)** The part dimensions and each part are modeled by 40 layers. **(d)** A laser beam is projected on the power bed. One side of the projected area is condensed laser energy and the other side is divergent.

*3.2. Experiment simplification for maximum pore size analysis*

Our goal is porosity generation analysis during LPBF AM process with the manufacturing setup introduced in section 3.1. Pore generation inside the part closely influences part mechanical properties. It is necessary to analyze part pore distribution to control part quality. One of the most influential factors affecting part quality is the maximum pore size. It is critical to analyze maximum pore generation, since big pores undermine the part's mechanical properties, such as strength, strain and fatigue, further leading to a part breakage and failure [40-41,46-48]. However, in practical experiments, it is really expensive, time-consuming, and labor-intensive to analyze all parts on the plates to investigate the maximum pore generation. Therefore, experiment simplification is critical and meaningful. By using the S-acceleration method to simplify an experiment plan before actually conducting experiments, the experiment trials that need to be conducted are reduced by at least half and, at the same time, achieve the goal of effectively investigating maximum pore generation. Based on our previous research [5,12], part porosity is closely related to machine setting conditions and physical effects of laser-materials interaction, which are distributed across parts; therefore, machine settings and physical effect distributions are selected to construct a semantic space for part mutual similarity measurement.

Machine settings are expressed by features such as laser power, part polar angle, azimuth angle, plate row coordinate, and plate column coordinate. Physical effect distribution is expressed by sixteen features: the average, standard deviation, minimum and maximum values of energy density, vertical pressures, horizontal pressures, and absolute pressure. These features were generated by analyzing 242 samples in the original experiment design for training the S-acceleration model. To describe microstructure inside parts, we selected four main characteristics: maximum pore size, which is the radius of the pore with maximum volume; mean pore size, which is the radius of the pore with mean volume; median pore size, which is the radius of the pore with median volume; and pore spacing, which is the spatial distance of two centroids of two pores inside the part. In order to obtain the above information, we first conducted X-ray tomography microscopy on the parts; then, based on the X-ray tomography data, we analyzed the pore size and spatial distributions. Using the raw tomography data as shown in Figure 4, an in-house batch analysis routine, Tomography Reconstruction Analysis and Characterization Routines (TRACR), was used to identify and analyze the internal porosity using a series of image processing steps and statistical tools. Response parameters such as maximum pore diameter, mean pore diameter, median pore diameter, and median pore spacing were derived using TRACR.

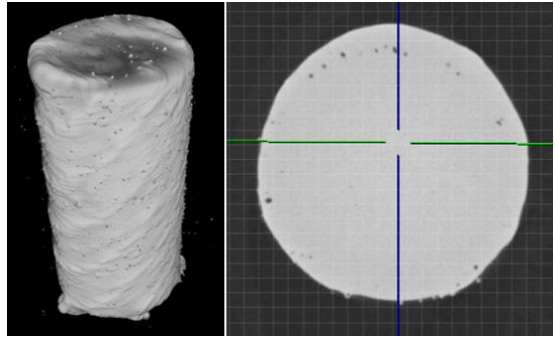

**Figure 4.** X-ray tomography of a compression sample. The topographical view (left) renders the sample's as-built surface, a cross-sectional view (right) reveals subsurface porosity.

### 3.3. Part semantics calculation

Generally, the semantic aspects for describing a part have three main aspects: machine setting conditions such as laser power, laser motion pattern, laser angle, part oblique angle, and part locations; physical effects such as laser energy density and laser pressure; and part defects such as maximum, median, and mean pore size distribution, pore spacing distribution, *etc.*

As a part of the manufacturing environment, machine settings decide laser working status, part spatial distribution, part pose placement, and powder supply speed and amount, all of which directly influence part performance as measured by various characteristics; therefore, the machine setting conditions for a part are good factors to be considered for part similarity measurement. Similar machine setting conditions will result in similar part performance. A typical laser-based machine setting is shown in Figure 3(a). Parts are built on a plate by a single laser or by multiple lasers. By scanning the powder bed with a laser, materials are melted in consecutive layers to form parts with different shapes. On each building plate in Figure 3(b), multiple parts are produced at the same time. Each part has a unique position and tilt angle. Laser angles and motion patterns are adjusted accordingly to produce the part.

The part properties are not only influenced by spatial system settings but also the accumulation of temporal physical effects. Typical physical effects during AM include energy density on part layers, directional pressure distributions on part layers, and so on. These physical effects have an impact on thermal conduction behaviors, microstructure formation, and the pore generations as well as mechanical properties. Therefore, it is critical to estimate physical effects using the temporal physics distribution during the manufacturing process to describe parts' mutual similarities. In the manufacturing process, the laser melts a building layer $j$ at a height $h_{ij}$ in the part $i$ as shown in Figure 3(c). The laser scanning angle $\theta$ is defined as the angle between the laser projection direction and the vertical direction. For laser energy distribution, half of the layer is condensed beam and the other half is divergent beam, shown in Figure 3(d). Given that a laser spot is tiny (about 0.07 $mm^2$) compared with part's cross-section layer area (3.14 $mm^2$), the average laser energy on the spot is selected to estimate energy density of laser beam in real-time roughly. The average energy density $e_{p_{ij}^k}$ on a specific point $p_{ij}^k$ in layer $j$ of part $i$ is calculated by Eq. (11), where $\theta_{p_{ij}^k}$ is the laser angle when point $p_{ij}^k$ is scanned, $A_0$ is the laser source area in vertical direction projection, and $w_0$ is the laser power (*e.g.*, $w_0$=160 $J/s$) for this part printing.

$$e_{p_{ij}^k} = e_0 \cos\theta_{p_{ij}^k} = w_0 \cos\theta_{p_{ij}^k}/A_0 \tag{11}$$

A laser beam contains energy and can exert a radiation force on the powder bed surface. To simplify our model, we assume the energy of a laser beam is fully absorbed by the powder layers. That means all the momentum of the laser beam is fully transferred to the radiation force. In a practical situation, the absorbed energy is a fixed proportion of a laser beam energy, and all the data will be normalized for training our model; therefore, this simplification does not significantly influence physical effect–based porosity prediction. Each photon has energy $E_0=hc/\lambda$ with momentum $p_0=E_0/c=h/\lambda$. The speed of light in a vacuum is represented by $c$, $h$ is Planck's constant, and $\lambda$ is the laser light wavelength. For each second, the total number of photons $M$ projected by a laser beam is calculated by the laser power $w_0$ divided by the photons' power $E_0$, as expressed by $M=w_0/E_0$. Therefore, the total radiation force $F_M$ exerted by a laser beam is calculated by the derivative value of $M$ photons' momentum $P$ and is shown in Eq. (12). The final radiation pressure $f$ exerted by a laser beam is calculated by the total force $F_M$ divided by the laser beam's projection area $A_1$, shown in Eq. (13). Based on the known laser angle $\theta$, pressure components along vertical and horizontal directions are calculated by Eq. (14) and Eq. (15), respectively.

$$F_M = \frac{dP}{dt} = \frac{d(Mp_0 t)}{dt} = Mp_0 = \frac{w_0}{hc} \tag{12}$$

$$f = \frac{F_M}{A_1} = \frac{w_0}{cs_0} \cos\theta_{p_{ij}^k} \tag{13}$$

$$f_v = \frac{w_0}{cs_0} \cos^2\theta_{p_{ij}^k} \tag{14}$$

$$f_h = \frac{w_0}{cs_0} \cos\theta_{p_{ij}^k} \sin\theta_{p_{ij}^k} \tag{15}$$

## 4. Results and Discussions

*4.1. Experiment simplification performance*

A sample experiment simplification with k-means unsupervised learning is shown in Figure 5. 242 parts are categorized into six clusters that are highlighted in a different color. In general, the parts in each cluster are symmetric about the central point of build plate, which indicates the model indeed captures the underlying physics effects of part since these parts are in symmetrical distribution in the spatial build plate. The simplified experiment plan uses 60% of the parts for case study, as shown in Figure 6(a). In comparison, in the evenly-select method, shown in green circles of Figure 6(b), parts are selected evenly by ignoring a fixed number of parts along each plate column. For example, when keeping 50% of the parts, a part will be selected and the next one will be ignored. In the simplified plan (green circles in Figure 6(a) denote selected parts for experiment), the parts in the clusters are intentionally selected and distributed on the build plate, showing good symmetry given the laser locations. Because these parts are produced by dual lasers (one worked on the upper half, and the other worked on the bottom half), good symmetry in both the whole plate and the half plates shows part selection effectiveness, which captures the influence of machine settings and manufacturing environment on the pore size distribution (part defect microstructure). Based on the captured part similarities, repeated parts that are relatively uninformative to build process-porosity models will be removed, and unique parts that are relatively informative to build process-porosity models will be retained.

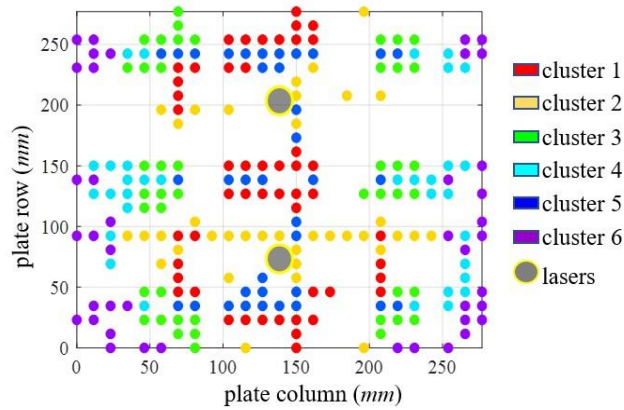

**Figure 5.** Visualization of part clusters using unsupervised learning algorithms. Each color denotes a cluster with respect to part mutual similarities. The machine settings of dual-laser position are shown in grey color and are set above of print plate plane at a height of 482.6 mm.

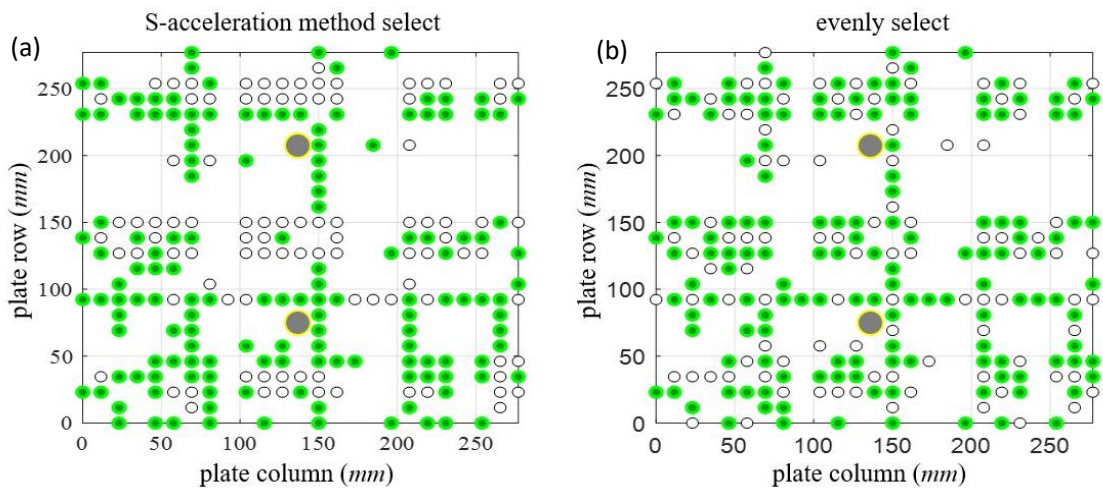

**Figure 6.** A sample comparison between part removals using S-acceleration method and "evenly select" method with experiment simplification degree of 0.6 (keep 60% of the original parts). Filled circles denote actual remaining parts during experiment simplification. Empty circles denote removed parts.

To show the unique advantages of S-acceleration methods in experiment simplification, the comparison was made among the original experiment plan with 242 experiment trials based on 242 parts ("ALL"), and the experiment plans made by S-acceleration method ("Similarity Selected"), Evenly select method ("Evenly Selected"), and randomly select method ("Randomly Selected") which selects experiment samples randomly, shown in Figure 7.

The evaluation criterion for method performance is the effectiveness and stability of part selection in maintaining the original pore size distribution. Because only when pore size distribution information against the process variations has been kept, the research conclusions related to process–porosity relations will be consistent in both original and simplified experiment plans. The original pore size distribution is important to keep for investigating both typical maximum pore generation and pore generation mechanisms. By using S-acceleration assisted parts selection (green bar), the pore distribution is consistent with the original pore distribution in the original experiment plan. As seen in Figure 7, in the popular maximum pore size range 36-119 µm, parts have been removed proportionally with stable bar height changes with S-acceleration similarity selection method, largely maintaining original pore size distribution; in the un-popular maximum pore size range 119-383 µm, 75% parts have also kept for the outliers of porosity investigation. By using an even selection method (blue bar), the original pore distribution has been changed with less stable performance in different ranges. For example, distributions in 36–42 µm, 63–70 µm and 77–91 µm have been over-sampled; while distributions in 49-56 µm, 70–77 µm, and 105–112 µm, have been under-sampled; in the large pore size range 119–383 µm, 62% parts have been kept, approximately keeping the outliers for porosity investigation. By using the random selection method (red bar), however, the original porosity distributions have been largely changed. For example, distributions in 36-42 µm, 56–70 µm and 139-167 µm have been over-sampled in the simplified plan; while distribution in 77–91 µm has been under-sampled; while in sparse ranges 174–383 µm, only 25% samples have been kept. This is because the random selection method is likely to ignore some uncommon pores, which account for small proportions in total experiment trials but is informative for building accurate process-porosity models. The good ability of S-acceleration method in maintaining the original porosity distribution shows the reliability of the S-acceleration method for experiment design.

To quantitatively analyze the performance of S-acceleration experiment design method, the method is conducted by different clustering algorithms and with different simplification degrees, as shown in Table 2. To avoid cherry-picking results, each running is performed ten times to get the average to measure the simplification performance. By comparing the simplified degree mean and STD relative with target simplification degrees, we can see that the similarity-based method is more accurate and stable than the random selection and even selection experiment plan. Using the similarity-based method, the absolute error 1% is achieved. The absolute error is defined as the difference between similarity-based simplified degree and target simplification degree. The error is relatively larger using the even selection method (absolute error 1-3%) and random selection method (absolute error 5-8%). The stability of method performance for different target simplification rates reveals the ability of S-acceleration experiment design method to keep the original porosity distribution for process-porosity relationships modeling. Stability is calculated by STD, given the expectation of the desired simplification degree. The similarity-based selection method is more stable (std 4-7%) than the even selection method (std 6-10%) and the random selection method (std 16-26%), showing better capability of the similarity-based method to maintain the original porosity distribution. The performance of another clustering method GMM is also performed for similarity-based part selection and exhibits similar results with k-means method. S-acceleration is also accurate and stable than the evenly select method and the randomly select method with a mean absolute error of 1% and STD 4–7%. This shows that desired simplification performances are brought by the semantic-based similarity selection model, instead of a specific unsupervised learning algorithm.

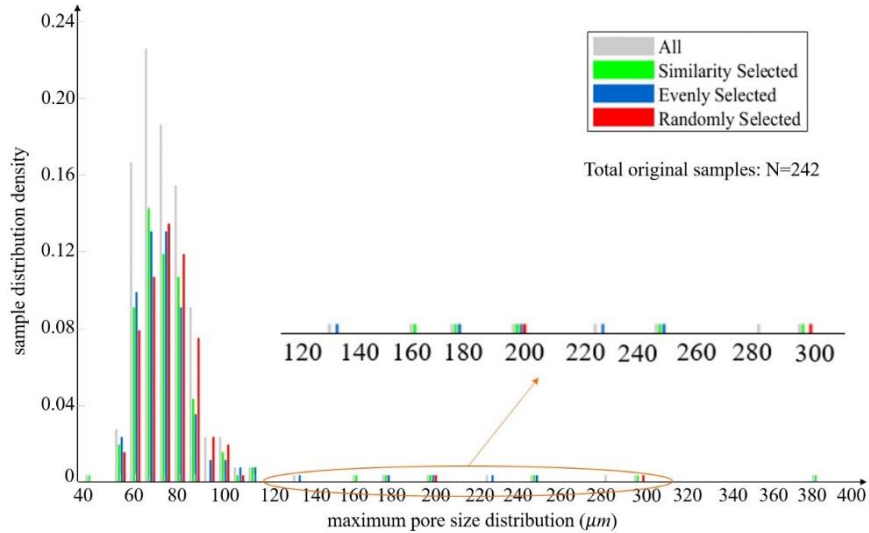

**Figure 7.** Maximum pore size distributions in simplified experiment plans designed with S-acceleration method (Similarity Selected), evenly selected and randomly selected methods.

**Table 2.** Performance comparisons with S-acceleration, evenly selected, and randomly selected method for different simplification degrees. Mean indicates the average of experiment simplification rate with three different methods, and STD means the corresponding standard deviation. The absolute error is defined as |simplified mean - target degree|.

|  | S-acceleration | | Evenly Select | | Randomly Select | |
| --- | --- | --- | --- | --- | --- | --- |
|  | Mean | STD | Mean | STD | Mean | STD |
| *deg* = 0.6 K-means | 0.59 | 0.07 | 0.57 | 0.10 | 0.65 | 0.16 |
| GMM | 0.59 | 0.07 | 0.57 | 0.10 | 0.65 | 0.16 |
| *deg* = 0.5 K-means | 0.49 | 0.06 | 0.48 | 0.06 | 0.57 | 0.26 |
| GMM | 0.49 | 0.04 | 0.48 | 0.06 | 0.58 | 0.26 |
| *deg* = 0.4 K-means | 0.39 | 0.05 | 0.40 | 0.08 | 0.47 | 0.25 |
| GMM | 0.41 | 0.07 | 0.41 | 0.08 | 0.48 | 0.25 |

## *4.2. Insights of S-acceleration method in experiment simplification*

To go beyond the black box of experiment simplification, the visualization of part selections and mutual part similarities within the cluster are shown in Figure 8, 9. As shown in Figure 8, each cluster emphasizes one or more different porosity ranges. Cluster 1 and 3 focused on removing parts in the range of 70-100 µm, while cluster 5 focused on the range of 56-70 *µm*, which are relatively close compared to similarities with other ranges. Different clusters may have common regions; for example, all of the 6 clusters focus on the range [63µm, 77µm]. Even for parts with the same porosity levels, pore generation mechanisms can be different. It is beneficial to have multiple clusters to emphasize different aspects of parts with the same porosity level for more detailed similarity measurements and more reasonable part removal. From Figure 9, it is obvious that parts in some clusters, such as clusters 1 and 3 with averaged similarity higher than 0.8, are highly repetitive and therefore are with a higher part removal rate (higher than the average removal proportion 40%). Meanwhile, parts in some clusters, such as clusters 2 and 6 with average similarity lower than 0.5, are less likely to be repeated and therefore are with a lower part removal rate (lower than the average removal proportion 40%). The similarity-based method adaptively adjusts the proportion value of part selection from these clusters according to the likelihood of a part being repeated, improving the accuracy of the experiment simplification plan.

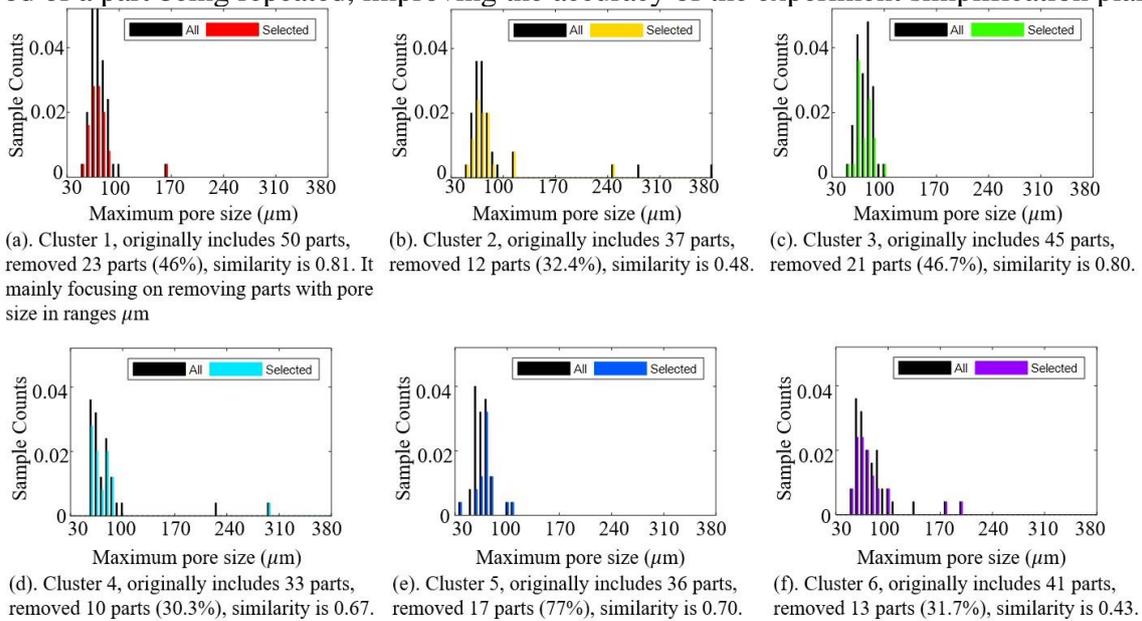

(a). Cluster 1, originally includes 50 parts, removed 23 parts (46%), similarity is 0.81. It mainly focusing on removing parts with pore size in ranges *µm*

(b). Cluster 2, originally includes 37 parts, removed 12 parts (32.4%), similarity is 0.48.

(c). Cluster 3, originally includes 45 parts, removed 21 parts (46.7%), similarity is 0.80.

(d). Cluster 4, originally includes 33 parts, removed 10 parts (30.3%), similarity is 0.67.

(e). Cluster 5, originally includes 36 parts, removed 17 parts (77%), similarity is 0.70.

(f). Cluster 6, originally includes 41 parts, removed 13 parts (31.7%), similarity is 0.43.

**Figure 8.** Maximum pore size distribution in each cluster. The black bar indicates the histogram of original parts cluster, and the color bar indicates selected parts for characterization as experiments simplification.

Note that, when considering both cluster spatial distributions (Figure 5), part removal ranges, and their average similarity scores (Figure 8), cluster numbers with high similarity score are approximated to the central region of print plate while other clusters with low similarity score are close to the plate borders. Results show that when the parts are manufactured in the central plate region, it is more likely to produce parts with high repetitiveness, whereas the parts produced in the marginal region are less likely repeated. In addition, the parts which are sparsely distributed with large laser-part variances are also less repeated. All of these can be explained that, when parts are received the appropriate amount of energy with desired energy variance, parts are more homogeneous, which will be mutually repeated and lead to obtaining similar process – property knowledge of metal AM; meanwhile, if parts do not receive appropriate laser energy, parts are more heterogeneous, which will be mutually distinguishable and informative for more rich knowledge gain during manufacturing.

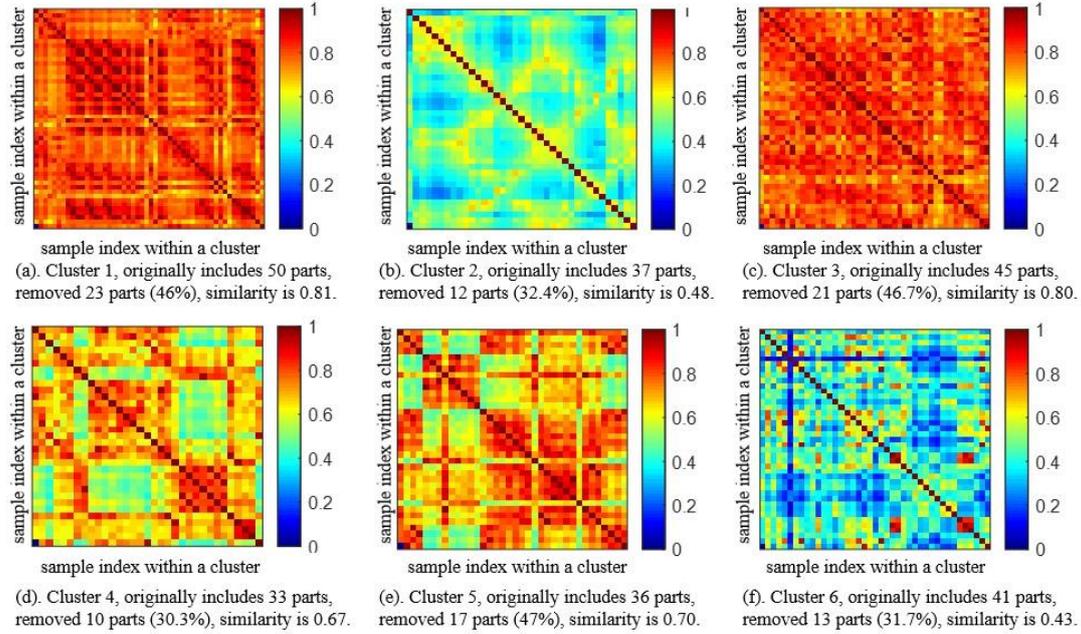

**Figure 9.** Mutual similarity visualization of parts inside a cluster. Bars indicate the value range of similarities.

*4.3. Uncertainty analysis*

Due to the S-acceleration method being supported by unsupervised learning algorithms, which are based on statistical learning, the statistical learning uncertainties are brought by using the S-acceleration experiment simplification. It is necessary to quantify the uncertainty influence on experiment simplification, further evaluating the S-acceleration method performance. With the same model parameters (simplification degree: 0.6, cluster number: 6, similarity threshold for detecting high-repeated clusters: 0.7, adaptive rate for increasing more part removal in high-repeated clusters: 20%), three times of the experiment simplifications have been done to generate three simplified experiment plans; by comparing the cluster difference in the three plans, simplification uncertainty is measured as shown in Figure 10. Four metrics have been used: similarity, n/clusterSize, std similarity and std n/clusterSize. The n/clusterSize denotes the ratio of part removal number within a cluster to the total number of the cluster, measuring what proportion of parts have been removed from a given cluster in different experiment plans; similarity denotes the average similarity scores of parts within a given cluster in different experiment plans, measuring the repeating degree of parts within a cluster; std n/clusterSize denotes the standard derivation of n/clusterSize, measuring the stability of the part selection in generating a simplification plan; std similarity denotes the standard derivation of similarity, measuring the stability of similarity score calculation in different experiment plans. As shown in Figure 10, when the average similarity score is higher than 0.7 (cluster 1, 3 and 5), more parts inside the cluster will be removed with a n/clusterSize more than 45% and a std n/clusterSize lower about 0.02; when average similarity score is lower than 0.7 (cluster 2, 4, 6), fewer parts inside a cluster will be removed with a n/clusterSize average 33.3% and an std n/clusterSize lower than 0.1. For all the clusters, the similarity variance is lower than 0.03. The small variances in both part removal proportion and similarity have proven that S-acceleration method is stable in simplifying experiments with smaller uncertainty.

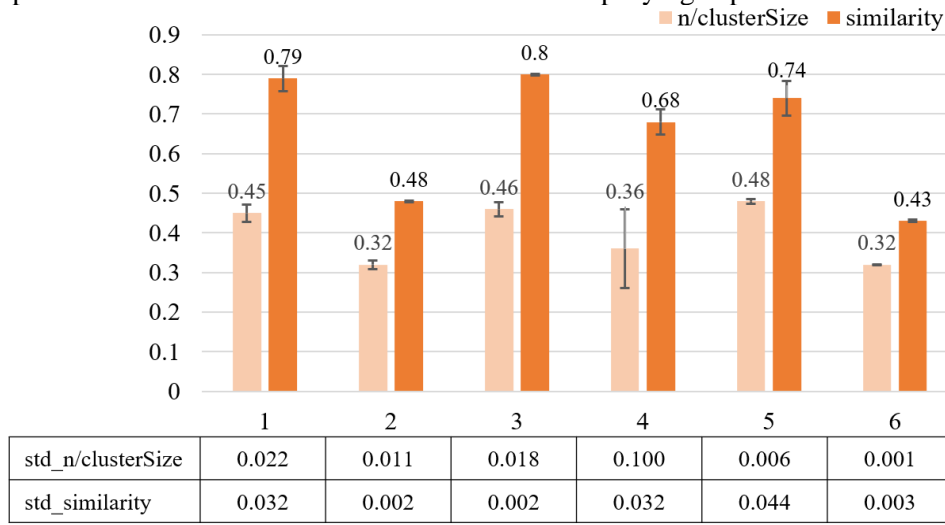

**Figure 10.** Uncertainty analysis during unsupervised part analysis when the simplification degree is 0.6. All the values were calculated by averaging the results from three times of metric calculations.

## 5. Conclusions

In this paper, we have developed a similarity-based S-acceleration method to simplify experiment design. A plate of parts is firstly printed for experimental data collection. To verify if the simplified experiments' parts porosity distribution is kept as the original full plate experiments, the S-acceleration method is applied on the data with different simplification rates to prove the feasibility of the framework. In practice, people can first design as many full factors experiments in regard to different machine settings and part pose parameters. With the help of our developed semantics part similarity metrics, the parts' mutual similarity can be quantified. The repeated parts are removed with a desired simplification rate without sacrificing knowledge obtained for process-porosity defects relationships. Thus, the total number of printing and part characterization is reduced to save time and materials resources. The s-acceleration experiment design framework is performed in three steps. Step one is part semantic similarity measurement, in which mutual part similarities are measured by using part semantic aspects, such as part machine setting conditions, part physical effect distributions and part microstructural defects. Step two is unsupervised learning of part clusters, in which parts with similar characters are grouped together. Step three is goal-guided experiment design, in which the parts with similar characteristics based on goal-desired simplification rate are removed from experiment trials for experiment simplification. We conducted an Inconel 718 pin part printing experiment on an LPBF concept laser M2 machine for relations between machine setting parameters and porosity defects property. Three levels of experiment simplification degrees (0.4, 0.5, 0.6) are defined for case study. The S-acceleration method has proved to be effective (mean error 1%, STD 4-7%) compared to even selection method (mean error 1-3%, STD 6-10%) and random selection method (mean error 5-8%, STD 16–26%) in both reducing experiment trials and maintaining the original characteristic value distribution. In the testing of the S-acceleration method on two unsupervised learning algorithms, k-means and the Gaussian mixture model show the effectiveness of S-acceleration experiment simplification on the real practice experiment design. In the future, this close-loop and one-shot-learning S-acceleration method will be improved to be a closed-loop experiment simplification method, which could involve human intelligence in simplification strategy making and evaluation, to improve the autonomous and executable of the S-acceleration experiment design method.

**Acknowledgments**

Use of instruments and characterization tools is supported by The Alliance for the Development of Additive Processing Technologies (ADAPT) Center at Colorado School of Mines. Thanks for valuable suggestions and experiment assistance from Dr. Aaron Stebner, Dr. Branden B. Kappes, Mr. Henry Geerlings, and Mr. Senthamilaruvi Moorthy. The partial support for Sen Liu at University of Louisiana at Lafayette is from the Research Competitiveness Subprogram (RCS) of the Board of Regents Support Fund (BoRSF) and Louisiana Transportation Research Center TIRE program.

**Conflicts of interest**

The authors declared that we have no conflicts of interest to this work.

**Authors' contribution**

Rui Liu and Sen Liu: formal analysis and writing, contributed equally. Xiaoli Zhang: supervision.